\documentstyle[seceq,twocolumn,epsf]{jpsj}
\def\runtitle{
 Fluctuation Effects
on the Quadrupolar Ordering
in Magnetic Field
}
\def\runauthor{Noboru {\sc Fukushima} and Yoshio \sc{Kuramoto}}
\title{
 Fluctuation Effects
on the Quadrupolar Ordering
in Magnetic Field
}
\author{Noboru {\sc Fukushima}\footnote{e-mail address :
 fukusima@cmpt01.phys.tohoku.ac.jp} and Yoshio \sc{Kuramoto}}
\inst{
Department of Physics, Tohoku University, Sendai 980-77, Japan
}

\recdate{ }
\abst{
Effects of magnetic field on the quadrupolar ordering
are investigated with inclusion of fluctuation of order parameters.
For the simplest model with the nearest-neighbor quadrupolar
interaction,
the transition temperature and the specific heat are derived by the
use of the recently proposed effective medium theory.
It is shown that
magnetic field $H$ has two competing effects
on the quadrupolar ordering;
one is to encourage the ordering by
suppressing the fluctuation
among different components of order parameters,
and the other is to block the ordering as in antiferromagnets.
The former is found to be of order $H^2$
and the latter of order $H^4$.
Hence the fluctuation
is suppressed for weak fields,
and the transition temperature
increases with magnetic field.
The fluctuation effect is so strong that the entropy released at the
quadrupolar ordering is only about half of the full value $\ln 4$ even
without the Kondo effect.
}
\kword{
quadrupole moment, CeB$_6$, TmTe, spherical model, effective medium,
 phase diagram, specific heat, Heisenberg model
}

\begin{document}
    \sloppy
    \maketitle

\section{Introduction}

Some compounds with orbital degeneracy show ordering of quadrupole
moments without magnetic order.
A typical example is CeB$_6$ which also shows the Kondo effect with
the Kondo temperature
$T_{\rm K}\sim 1 \rm K$\cite{Tk}.
The Ce$^{3+}$ multiplet (4f$^1$, $J=5/2, S=1/2, L=3$)
is split under the cubic crystal field
 into the ground-state quartet $\Gamma_8$ and
an excited-state doublet $\Gamma_7$  separated by about
540K.\cite{split}
It has a transition from the phase I (paramagnetic phase)
 to the phase II at 3.3K under zero field.
Neutron\cite{neutron} and NMR\cite{NMR1,NMR2} experiments
 show that
electric quadrupole moments are ordered
 with a periodic pattern  in the phase II.
Hence it is also called the antiferro-quadrupolar phase (AFQ).
With further decrease of temperature, another transition occurs at
2.4K
to the phase III, which accompanies the magnetic ordering
with a complicated periodic pattern.
The phase boundary between the phases I and II
 shows unusual dependence on magnetic field;
 the transition temperature $T_{\rm Q}(H)$ increases
 with magnetic field $H$.
Namely, while a magnetic field will ordinarily destroy an ordered
phase,
 it assists the quadrupolar ordering in CeB$_6$.
The phase boundary shifts to lower temperatures
in the dilute alloy Ce$_{1-x}$La$_x$B$_6$.\cite{Goto,Sera}
The AFQ is found also in TmTe,\cite{Matsumura} which is insulating.
Though its phase diagram has something similar to that of CeB$_6$,
 $T_{\rm Q}(H)$ shows a maximum as a function of
 magnetic field in contrast to that of CeB$_6$.
In the latter, $T_{\rm Q}(H)$ continues to increase up to
 the maximum field measurable at present.

One of the keys to understanding
 the transition between the phases I and II should be
 the result of specific-heat measurement,\cite{Matsumura,hinetsu}
 which indicates large fluctuation there.
Although the peak at the AFQ transition is
 small in zero magnetic field,
 it grows larger with increasing magnetic field.
The small peak suggests that
 only a small change of entropy is involved  at the transition,
and that a large amount of short-range order should exist even above
$T_{\rm Q}$.
In other words, the long-range order is
 disturbed by some large fluctuation.
Therefore, the experimental result of specific heat
 suggests that the large fluctuation should be
 suppressed by a magnetic field.
The purpose of this paper is to clarify
 how a magnetic field affects the fluctuation
  at the quadrupolar ordering.
We take account of fluctuation in our calculation
with use of an effective medium theory for quantum spins and
multipoles.\cite{KF}

Several models for the quadrupolar ordering in CeB$_6$ have been
proposed.
Hanzawa and Kasuya\cite{Hanzawa}
 assumed that the ground state is the doublet $\Gamma_7$,
 and considered the quadrupolar interaction among Ce ions.
In this model, large off-diagonal elements
 of quadrupole-moment operators
 between $\Gamma_7$ and $\Gamma_8$
 become important as magnetic field increases.
Thus, $T_{\rm Q}(H)$ increases with $H$ in the mean-field
approximation.
Although this model should be effective for the system with
the $\Gamma_7$ ground state,
 it was shown later by experiment that
$\Gamma_8$ is the ground state and $\Gamma_7$
 is an excited state.\cite{split}
Another model proposed by Ohkawa\cite{Ohkawa1,Ohkawa2}
 has the symmetric form between the spin doublet and the orbital
doublet represented by pseudo spins.
The Hamiltonian includes an intersite interaction which is biquadratic
in spin and quadrupole moments.
This interaction enhances the quadrupolar interaction with increasing
$H$.
Hence $T_{\rm Q}(H)$ increases in the mean-field approximation
 also in this model.
Shiina {\it et.al.}~\cite{Shiina}
 recently rewrote Ohkawa's model in the form including
magnetic octupoles and dipoles,
and solved the resultant model by the mean-field approximation.
Finally, Uimin and the present authors
 treated the effect of fluctuation.\cite{UKF,Uimin}
They approximated the pseudo-spins by classical spins
and applied the spherical model which can include fluctuation in a
simple manner.
As a result, the transition temperature increased
 with magnetic field in contrast to the result of
 the mean-field approximation.

In this paper we accomplish substantial improvement over the theory of
ref.\citen{UKF}.
The fluctuation emphasized there is due to the competition
 among order parameters with  different wave numbers $\mib k$.
The intersite interaction was taken to be
the electrostatic one between the $\Gamma_3$-type quadrupole moments.
Strength of the quadrupolar interaction at $\mib k=(\pi\pi\pi)$ is
almost the same as that at $\mib k=(\pi\pi 0)$
in the case of the electrostatic interaction.
Competing fluctuations are suppressed by a magnetic field.
However, as will be discussed in detail later, what is essential
of the magnetic field is to suppress competition among {\it
components} of fluctuating order parameters, namely, ${O_2}^0$ and
${O_2}^2$.
The presence of competition among order parameters with different wave
numbers helps the suppression effect mentioned above.
The value of ${\rm d} T_{\rm Q}(H)/{\rm d} H$ for the electrostatic interaction 
 is about twice as large as that for the nearest-neighbor one.
Although this enhancement is substantial quantitatively, we regard
that the specific form of interaction is not essential for qualitative
understanding.
Hence we adopt the nearest neighbor interaction for simplicity of
calculation
in this paper.
Further, we do not restrict the quadrupolar interaction
to that of the $\Gamma_3$-type moments.

\section{Model for Quadrupolar Interaction}

\subsection{Model}

Our model corresponds to a localized electron system
 which has interaction between quadrupole moments only.
Possible targets of the model are not only to CeB$_6$ but also
related quadrupolar systems under the conditions to be described
below.
An insulating compound TmTe has a phase diagram similar to CeB$_6$.
If the increase of transition temperature in these two
materials
 comes from a common mechanism,
the interaction with conduction electrons
should not be essential.
Hence we do not consider this interaction explicitly.

Cerium sites in CeB$_6$ constitute the simple cubic lattice.
At temperatures near $T_{\rm Q}$,
 the excited CEF states, which are separated by
 about 540K in CeB$_6$,\cite{split}
 have little influence.
Therefore we deal with the ground-state quartet $\Gamma_8$ only.
We represent here the basis set of the $\Gamma_8$ states
 with use of the $|J^z\rangle$ basis,
\begin{equation}
|\psi_{1\pm}\rangle=\sqrt{\frac{5}{6}}\left| \pm \frac{5}{2}
\right\rangle
         +\sqrt{\frac{1}{6}}\left| \mp \frac{3}{2} \right\rangle
\;,\ \;\;
|\psi_{2\pm}\rangle=\left| \pm \frac{1}{2} \right\rangle.
\end{equation}
The Hamiltonian for our model is composed of
the quadrupolar interaction term ${\cal H}_{\rm int}$
 and the Zeeman term  ${\cal H}_{\rm Z}$:
\begin{equation}
{\cal H}  =  {\cal H}_{\rm int} + \sum_i {\cal H}_{\rm Z}(i),
\end{equation}
\begin{equation}
\begin{array}{l}\displaystyle
{\cal H}_{\rm int} = J_{\Gamma 3} \sum_{\langle ij\rangle}
        \left[ {O_2}^0(i) {O_2}^0(j) + {O_2}^2(i) {O_2}^2(j) \right]
\\ \displaystyle
 +J_{\Gamma 5}\! \sum_{\langle ij\rangle}
   \left[ O_{yz}(i) O_{yz}(j) \! + \! O_{zx}(i) O_{zx}(j)
                                 \! +\! O_{xy}(i) O_{xy}(j)\right],
\end{array}
\end{equation}
\begin{equation}
{\cal H}_{\rm Z}(i) = - \frac67 \mu_{\rm B} {\mib J}(i) \cdot
{\mib H},
\end{equation}
where $6/7$ in the Zeeman term is the $g$ factor.
The summation in ${\cal H}_{\rm int}$
is performed over all the nearest neighbor pairs.
Quadrupole-moment operators at each site are given as
\begin{equation}
\label{eq:Ogamma3}
\begin{array}{c}\displaystyle
 {O_2}^0 =  \frac1{\sqrt{3}} [3 (J^z)^2 - J(J+1)],\\ \displaystyle
 {O_2}^2 =  (J^x)^2 - (J^y)^2,
\end{array}
\end{equation}
\begin{equation}
\label{eq:Ogamma5}
\begin{array}{c}
 O_{yz}  =  J^y J^z + J^z J^y, \\ \displaystyle
 O_{zx}  =  J^z J^x + J^x J^z,\\ \displaystyle
 O_{xy}  =  J^x J^y + J^y J^x.\\ \displaystyle
\end{array}
\end{equation}

The interaction constant
$J_{\Gamma 3}$ is different from $J_{\Gamma 5}$ in general.
We take three types of interaction constants:
(i) $\Gamma_3$-type interaction  $J_{\Gamma 3} > 0, J_{\Gamma 5}=0$;
(ii) $\Gamma_5$-type interaction   $J_{\Gamma 3}=0, J_{\Gamma
5}>0$;
(iii) $(\Gamma_3+\Gamma_5)$-type interaction
 $16 J_{\Gamma 3}=J_{\Gamma5}>0$.
In the case (iii), interaction constants are taken so that
every ordering of five types of quadrupole moments
has the same transition temperature in zero magnetic field.
This becomes clear in the pseudo-spin representation
described in the next subsection.
It is expected that the effect of fluctuation is the largest in the
case (iii).

This model is  the simplest one for the quadrupolar ordering
in CeB$_6$.
However, in the literature the increase of
 the transition temperature
 has never been discussed
 within this model.
This is probably because the result of the mean-field approximation
does not
 show the increasing transition temperature.
We treat this model beyond the mean-field approximation
to discuss effects of the fluctuation.

\subsection{Pseudo-spin representation}

It is convenient for calculation of $T_{\rm Q}(H)$ to use the
pseudo-spin representation.
When the Hilbert space is limited to the $\Gamma_8$ states,
 all operators are represented by these pseudo-spins
 because all the matrix elements can be represented by those of
pseudo-spins.
The pseudo-spin operators are defined by
\begin{equation}
\begin{array}{cl}
 \tau^z |\psi_{1\pm}\rangle =   |\psi_{1\pm}\rangle,&
 \tau^z |\psi_{2\pm}\rangle = - |\psi_{2\pm}\rangle,\\
 \tau^+ |\psi_{2\pm}\rangle =   |\psi_{1\pm}\rangle,&
 \tau^- |\psi_{1\pm}\rangle =   |\psi_{2\pm}\rangle,\\
 \sigma^z |\psi_{\alpha +}\rangle = |\psi_{\alpha +}\rangle,&
 \sigma^z |\psi_{\alpha -}\rangle =-|\psi_{\alpha -}\rangle,\\
 \sigma^+ |\psi_{\alpha -}\rangle = |\psi_{\alpha +}\rangle,&
 \sigma^- |\psi_{\alpha +}\rangle = |\psi_{\alpha -}\rangle,\\
 & \hspace{2.5em}(\alpha=1,2).
\end{array}
\end{equation}
Note that this definition is different
 from that in refs.\citen{Ohkawa1,Shiina} and \citen{UKF}
where the positive eigenvalue of the pseudo-spins is set to 1/2.
It is set to unity here  to remove the constant 1/2 from various
expressions.

In this representation, quadrupole-moment operators are given by
\begin{equation}
 {O_2}^0 = \frac8{\sqrt{3}}\tau^z,\;\;
 {O_2}^2 = \frac8{\sqrt{3}}\tau^x,
\end{equation}
\begin{equation}
 O_{yz} = \frac2{\sqrt{3}}\tau^y\sigma^x,\;\;
 O_{zx} = \frac2{\sqrt{3}}\tau^y\sigma^y,\;\;
 O_{xy} = \frac2{\sqrt{3}}\tau^y\sigma^z.
\end{equation}
The interaction Hamiltonian is rewritten as
\begin{eqnarray}
\label{eq:Hint-psd}
{\cal H}_{\rm int} & = & J_{\Gamma 3}' \sum_{\langle ij\rangle}
        \left[ \tau^z(i)\tau^z(j) +  \tau^x(i)\tau^x(j)  \right]
\nonumber \\
 & & +  J_{\Gamma 5}' \sum_{\langle ij\rangle}
   \tau^y(i)\tau^y(j) {\mib \sigma}(i) \cdot {\mib \sigma}(j)   ,
\end{eqnarray}
with $J_{\Gamma 3}'=J_{\Gamma 3}\times 64/3$ and
$J_{\Gamma 5}'=J_{\Gamma 5}\times 4/3$.
Hence $J_{\Gamma 3}'=J_{\Gamma5}'$ is satisfied
 for the $(\Gamma_3+\Gamma_5)$-type interaction.
The Zeeman term is rewritten as
\begin{equation}
{\cal H}_{\rm Z}(i)  =  - \sum_{\alpha=x,y,z}
 \mu_{\rm B} \left(1+\frac47 T^\alpha(i)\right)\sigma^\alpha(i)
H^\alpha ,
\end{equation}
where $T^\alpha=
 \frac1{\sqrt{3}}\{3(J^\alpha)^2-J(J+1)\}\times \frac{\sqrt{3}}{8}$
is given in the pseudo-spin representation by
\begin{equation}
 T^x=-\frac12\tau^z+\frac{\sqrt{3}}{2}\tau^x ,\;\;
 T^y=-\frac12\tau^z-\frac{\sqrt{3}}{2}\tau^x ,\;\;
 T^z= \tau^z .
\end{equation}
Intuitively,
$(1+4/7 \; T^\alpha)$ represents the magnitude $|m|$
of the magnetic moment with $3/7 \leq |m| \leq 11/7$
 and $\sigma^\alpha$ represents the orientation of it.

\section{Static Approximation for the Effective Medium}
\label{Method}

We recently formulated a dynamical effective medium theory for
 quantum spins and multipoles. \cite{KF}
The theory is a generalization of
 the spherical model approximation for the Ising
model.\cite{Brout60,Brout}
It takes account of fluctuation up to $O(1/z_n)$,
 where $z_n$ is the number of interacting neighbors.
We use the static approximation (SA) which neglects the dynamical
nature of fluctuations.
Although it is difficult to estimate the accuracy of the SA in the
temperature range of the quadrupole order, we use the SA as our first
step to study the importance of fluctuations.
Since derivation of this method is given in ref.\citen{KF},
we explain here some characteristics of this method
and details of the calculation.

%
In the mean-field approximation (MFA),
 one solves an effective single site problem,
 where
 an operator for interacting neighbors is replaced by an effective
field.
 That field is zero in the disordered phase.
On the other hand,
 the effective field in the SA is
 distributed with Gaussian distribution around a mean value
 even in the disordered phase.
This fluctuating field represents the local field
 created by the short-range order.

In order to explain the method concisely,
we represent quadrupole-moment operators as $O^\alpha$ symbolically.
We use the irreducible tensor operators for $O^\alpha$
to obtain scalar equations from a matrix equation.\cite{KF}
For $\mib H \| (001)$, $O^\alpha$ refers to one of the operators
in eq.(\ref{eq:Ogamma3}) and/or eq.(\ref{eq:Ogamma5}).
For $\mib H \| (111)$,  eq.(\ref{eq:Ogamma3}) and/or
$(-O_{yz}-O_{zx}+2O_{xy})/\sqrt{6}$,
$(O_{yz}-O_{zx})/\sqrt{2}$ and
$(O_{yz}+O_{zx}+O_{xy})/\sqrt{3}$.
Then, the Hamiltonian for our model can be expressed as
\begin{equation}
{\cal H} = - \frac12\sum_{ij}\sum_{\alpha=1}^n J_{ij}^\alpha
O^\alpha(i) O^\alpha(j)
+ \sum_i {\cal H}_{\rm Z}(i),
\end{equation}
where $n$ refers to the number of fluctuating components and
is equal to 2, 3 and 5
for the $\Gamma_3$-, the $\Gamma_5$- and the
$(\Gamma_3+\Gamma_5)$-type
 interactions respectively.

We write the mean field as
 $a^\alpha(i) = \sum_{j} J_{ij}^\alpha \langle O^\alpha (j) \rangle$
 and
 the fluctuating field around it as $\zeta^{\alpha}(i)$.
The effective single-site Hamiltonian ${\cal H}_1$
and the partition function $Z_1$ is given by
\begin{equation}
{\cal H}_1 =  \sum_{\alpha=1}^n
  \{- a^{\alpha} O^{\alpha}
 - \zeta^{\alpha}( O^{\alpha} -\langle O^{\alpha} \rangle ) \}
 + {\cal H}_{\rm Z} ,
\end{equation}
\begin{equation}
Z_1\!=\!\! \prod_{\alpha=1}^n \left\{ \int_{-\infty}^{\infty}
 \frac{\beta \: {\rm d}\zeta^{\alpha}}{\sqrt{2 \pi \beta \tilde{J}^\alpha}}
 \: \exp[-\beta \frac{(\zeta^{\alpha})^2}{2 \tilde{J}^\alpha}]
                        \right\} \!
\mbox{Tr}  \exp[ -\beta {\cal H}_1] ,
\end{equation}
where the site index $i$ is omitted.
The term
$ \zeta^{\alpha}\langle O^{\alpha} \rangle$ in the Hamiltonian
 plays a role to enforce $\langle\zeta^{\alpha}\rangle=0$.
The variance  $\tilde{J}^\alpha$ is determined
 by the self-consistency condition
\begin{equation}
\label{eq:sumrule}
\chi_{\rm L}^{\alpha}=\int \frac{{\rm d}^D q}{(2\pi)^D}
 \frac{\chi_{\rm L}^{\alpha}}
 {1-(J^\alpha_{\mibs q}-\tilde{J}^\alpha)\chi_{\rm L}^{\alpha}} ,
\end{equation}
where $D$ is the spatial dimensions ($D=3$ for the present case),
$\chi_{\rm L}$ is the local (strain) susceptibility
calculated with ${\cal H}_1$,
and $J^\alpha_{\mibs q}$ is the Fourier transform of the interaction:
$J^\alpha_{\mibs q}=\sum_j J^\alpha_{j0}\exp [-i {\mib q \cdot (\mib
R_{j}-\mib R_{0})}]$.
The lattice constant is set to unity.
This equation requires the consistency
 between the local susceptibility and its another expression
 using the inverse Fourier transform:
$\chi_{\rm L}=N^{-1}\sum_{\mibs q} \chi_{\mibs q}$.
Note that this consistency is not satisfied in the MFA
 for a finite dimensional system.
The integral can be performed analytically for the nearest neighbor
interaction $J^\alpha_{\mibs q}= -2 J^\alpha \sum_{l=1}^D \cos q_l$
 even in three dimensions. \cite{Joyce}

%
The SA still keeps the
 noncommuting character of quantum operators.
In contrast, our previous theory for CeB$_6$\cite{UKF}
replaces the pseudo-spin vectors composed of the Pauli matrices
by classical vectors.
Further improvement of
the present theory is that
the renormalization is performed differently for each component
 $\tilde{J}^\alpha$ when the model has a low symmetry.
The spherical model for classical spins used in ref.\citen{UKF}
cannot deal with the anisotropy in $\tilde{J}^\alpha$
even if the symmetry is lowered by a magnetic field.
These theories coincide with each other
 in the limit of infinite temperature.

At high temperatures, the variance $\tilde{J}^\alpha$
 in Gaussian distribution of the effective field is almost zero, and
 therefore it has little deviation from the MFA.
As the temperature decreases,
the short-range order makes $\tilde{J}^\alpha$ larger.
In other words, the distribution of fluctuations gets broader.
Finally it reaches the critical variance $\tilde{J}^{\rm cr}$
 at the critical temperature, and the transition occurs;
 the susceptibility
 for a certain component $\lambda$
(or for multiple components simultaneously) diverges
 with the wave number ${\mib q}={\mib Q}$
 at which $J^\lambda_{\mibs q}$ takes the maximum
 $J^\lambda_{\mibs Q}$.
Hence, we obtain
\begin{equation}
\label{eq:tc}
\chi_{\rm L}^{\lambda{\rm cr}}=\int \frac{{\rm d}^D q}{(2\pi)^D}
 \frac1{J^\lambda_{\mibs Q}-J^\lambda_{\mibs q}}
\end{equation}
and
$\tilde{J}^{\lambda \rm cr}
=J^\lambda_{\mibs Q}-(\chi_{\rm L}^{\lambda{\rm cr}})^{-1}$.

As the number $D$ of spatial dimensions goes to infinity,
 the critical variance $\tilde{J}^{\rm cr}$ approaches zero.
In short, the SA approaches the MFA.
This is consistent with the nature of the theory;
the original theory for the SA is
 the next leading order theory  with respect to the inverse of
dimensions,
while the MFA is the leading order theory.
Small $\tilde{J}^{\rm cr}$ leads to
a large transition temperature
 because the effect of fluctuation is small.
On the contrary, when the number $n$ of
 fluctuating components increases,
 the critical temperature becomes smaller.
The reason is that
the local susceptibility $\chi_{\rm L}$ decreases
 with increasing $n$ as shown below.

To perform the multiple Gaussian integral,
 the polar coordinate in the $\zeta$ space is convenient.
The reason is that in zero field some integrands of angular integral
  are constants by symmetry and
 remain smooth even in weak magnetic fields.
For example, in terms of the polar coordinate for
the $(\Gamma_3+\Gamma_5)$-type interaction, we obtain
\begin{equation}
\begin{array}{l} \displaystyle
 \prod_{\alpha=1}^5 \int_{-\infty}^{\infty}
 \frac{\beta \: {\rm d}\zeta^{\alpha}}{\sqrt{2 \pi \beta \tilde{J}^\alpha}}
 \: \exp[-\beta \frac{(\zeta^{\alpha})^2}{2 \tilde{J}^\alpha}]
\\ \displaystyle \hspace{1em}
= \frac1{(\sqrt{2 \pi})^5}
 \int {\rm d}\Omega_{\theta} \int_0^{\infty} \zeta_r^4 {\rm d}\zeta_r
 \exp[-\frac{\zeta_r^2}{2}],
\end{array}
\end{equation}
where $\zeta_r=\sqrt{\sum_\alpha \beta(\zeta^\alpha)^2 /
\tilde{J}^\alpha}$ and
$\Omega_{\theta}$ refers to the generalized solid angle.
To perform the integral on the unit hypersphere in five dimensions,
 we introduce the variables $\theta_a, \theta_b \cdots$ as follows:
\begin{equation}
\begin{array}{rcl}
\zeta_1 & = & \sqrt{\tilde{J}_1/\beta}\;\zeta_r \sin\theta_a
\sin\theta_b \sin\theta_c \sin\theta_d, \\
\zeta_2 & = & \sqrt{\tilde{J}_2/\beta}\;\zeta_r \sin\theta_a
\sin\theta_b \sin\theta_c \cos\theta_d, \\
\zeta_3 & = & \sqrt{\tilde{J}_3/\beta}\;\zeta_r \sin\theta_a
\sin\theta_b \cos\theta_c, \\
\zeta_4 & = & \sqrt{\tilde{J}_4/\beta}\;\zeta_r \sin\theta_a
\cos\theta_b, \\
\zeta_5 & = & \sqrt{\tilde{J}_5/\beta}\;\zeta_r \cos\theta_a,
\end{array}
\end{equation}
\begin{equation}
\label{eq:hyperint}
{\rm d}\Omega_{\theta}=
 \sin^3\theta_a \sin^2\theta_b \sin\theta_c
{\rm d}\theta_a {\rm d}\theta_b {\rm d}\theta_c {\rm d}\theta_d ,
\end{equation}
with $\theta_a, \theta_b, \theta_c \in [0,\pi]$
and $\theta_d \in [0,2\pi)$.

%
In zero field, our model has a high symmetry so that
$\chi_{\rm L}$ and $\tilde{J}$ do not depend on $\alpha$.
Then,
$\sum_{\alpha=1}^n\chi_{\rm L}^{\alpha}/n$ has
 the spherical symmetry in the $\zeta$ space
 and is equal to $\chi_{\rm L}$.
We obtain
\begin{eqnarray}
\chi_{\rm L} & = & \frac{\beta}{n} \left(1+  (n-1)\frac
 {\langle \sinh(\beta\zeta)/(\beta\zeta)
  \rangle_\zeta}
   {\langle\cosh(\beta\zeta)\rangle_\zeta}  \right) \\
\label{eq:chiL}
& = & \frac{\beta}{n} \left(1+  (n-1)\frac
  {F(n/2,3/2;\beta\tilde{J}/2)}
  {F(n/2,1/2;\beta\tilde{J}/2)} \right)
\end{eqnarray}
where $\langle \cdots \rangle_\zeta$ stands for
 $ \int_0^{\infty}{\rm d}\zeta \cdots
 \zeta^{n-1} \exp[-\beta\zeta^2/(2 \tilde{J})]$
and $F(\alpha,\gamma;z)$ is Kummer's confluent hypergeometric
function.
In fact, this equation is common to the results of the SA for the
Heisenberg model($n=3$), the XY model($n=2$) and the Ising
model($n=1$)
in zero field.
Let us take such spin models to see the dependence on $n$.
The first term 1 in the bracket of eq.(\ref{eq:chiL})
represents contribution from a fluctuating field
 parallel to the spin
 and the second term with factor $n-1$ perpendicular to it.
We note that the quotient
$F(\cdots)/F(\cdots)$ in the second term of eq.(\ref{eq:chiL})
is less than unity except for $\beta\tilde{J}=0$.
The quotient is reduced to a simple form $(1+\beta\tilde{J})^{-1}$ for
$n=3$,
and $(3+\beta\tilde{J})(3+6\beta\tilde{J}+\beta^2\tilde{J}^2)^{-1}$
for $n=5$.
For $n=2$, it takes a rather complicated form
 including the incomplete gamma function.
As $n$ becomes larger, the number of perpendicular components
increases
and thus the perpendicular part in eq.(\ref{eq:chiL}) becomes
dominant.
Hence it leads to the decrease of $\chi_{\rm L}$.
It can be shown by more detailed investigation that
$\chi_{\rm L}$
is a monotonically decreasing function of $n$
 at a fixed temperature.
Physically, the short-range order of the $z$ component disturbs
 orderings of $x$ and $y$ components.
This corresponds in the SA to the fact that
 the distribution of $\zeta^z$ makes
  $\chi_{\rm L}^x$ and $\chi_{\rm L}^y$ decrease
 and the orderings of $x$ and $y$ components become less favorable.
In \S\ref{sec:gauss}, we investigate the effect in more detail.
This effect
 becomes large as the number of fluctuating components increases.
We note that the Ising model with $n=1$ does not have
 such fluctuation effect.

Once a magnetic field is applied, the model has a lower symmetry and
each $\tilde{J}^\alpha$ can have different values.
As temperature decreases, one of them reaches the critical value
and satisfies eq.(\ref{eq:tc}).
Hence one must solve the set of nonlinear equations
 to determine the other $\tilde{J}^\alpha$'s and the critical
temperature.
We used the Newton method to solve the set of equations.
In solving it,
we took care of the fact
 that $(\chi_{\rm L}^{\alpha})^{-1}+\tilde{J}^\alpha$
 in the integral in eq.(\ref{eq:sumrule})
 is always larger than $J^\alpha_{\mibs Q}$.
At finite magnetic fields,
 the multiple Gaussian integral
  could not be performed analytically,
and therefore we did numerical integration.
For the $\Gamma_3$- and the $\Gamma_5$-type interactions,
the integral is no more than a double integral by symmetry,
and thus executable.
However, for the $(\Gamma_3+\Gamma_5)$-type interaction,
 the integral on the unit hypersphere
 represented by eq.(\ref{eq:hyperint})
 was too time-consuming to use the standard routine of numerical
integration.
Therefore, we approximated the integral
by summation over ten points
$(\pm1,0,0,0,0),(0,\pm1,0,0,0),\cdots,(0,0,0,0,\pm1)$
 on the unit hypersphere in five dimensions.
This method is exact in zero field.
We have checked that
the same approximation for
the $\Gamma_3$- and the $\Gamma_5$-type interactions
gives only a few percent of
deviation in the transition temperature
 from the results with use of the double integral.

For calculation of the specific heat,
we use the formula $C={\rm d} U/{\rm d} T$ where
the internal energy $U= \langle {\cal H} \rangle$
is differentiated numerically.
In this paper, we restrict the calculation only to
 the disordered phase.
Then,
expectation values of quadrupole moments $\langle O^\alpha(i)\rangle$
and dipole moments $\langle \mib M (i) \rangle$ do not depend on the
site
index $i$.
The internal energy is written within the SA as
\begin{eqnarray}
U & = & -   \frac T 2
\sum_{\mibs q}\sum_{\alpha=1}^n J_{\mibs q}^\alpha
 \chi_{\mibs q}^\alpha
-\frac12  \sum_{ij} J^\alpha_{ij} \langle O^\alpha \rangle^2
- N \mib H \cdot  \langle \mib M \rangle \nonumber \\ & & \\
& = &
 -N\left( \frac T 2
 \sum_{\alpha=1}^n \tilde{J}^\alpha \chi_{\rm L}^\alpha
+\frac12 J_{\mibs q=0}^\alpha \langle O^\alpha \rangle^2
+  \mib H  \cdot \langle \mib M \rangle \right), \nonumber \\
\label{eq:energy} & &
\end{eqnarray}
where $\chi_{\mibs q}^\alpha=\chi_{\rm L}^{\alpha}
 \{1-(J^\alpha_{\mibs q}-\tilde{J}^\alpha)\chi_{\rm
L}^{\alpha}\}^{-1}$ and the site index is omitted.
We used here the relation
$N^{-1}\sum_{\mibs q}
 J_{\mibs q}^\alpha \chi_{\mibs q}^\alpha
=\tilde{J}^\alpha\chi_{\rm L}^\alpha$,
which is derived from eq.(\ref{eq:sumrule}).
All of the quantities in eq.(\ref{eq:energy}) can be
calculated with the effective single-site Hamiltonian ${\cal H}_1$.

\section{Numerical Results}

\subsection{Quadrupolar ordering temperature}
In contrast to the MFA, the present effective medium theory with the
SA gives the result
that $T_{\rm Q}(H)$ increases  with $H$.
Figure 1 shows the phase diagrams obtained by the present theory
and those by the MFA for comparison.
The low temperature phase is the antiferro-quadrupolar phase.
The magnetic field is oriented along either (001) or (111).
As magnetic field increases, numerical integration of
 Gaussian distribution becomes difficult.
Hence the phase diagrams are plotted up to the maximum field
 we can calculate.

We take the interaction constant $J'$ as the unit of energy,
 where $J'$ refers to either $J_{\Gamma3}'$ or $J_{\Gamma5}'$
in eq.(\ref{eq:Hint-psd}).
Accordingly $J'/k_{\rm B}$ is taken to be the unit of temperature
and $J'/\mu_{\rm B}$ the unit of magnetic field.
In this unit, $T_{\rm Q}(H=0)$ in the MFA is 6 $(=z_n)$ irrespective
of the number of competing order parameters.
With inclusion of fluctuations we have the following relation:
$$
T_{\rm Q}^{\Gamma 3+\Gamma 5} < T_{\rm Q}^{\Gamma 5} < T_{\rm
Q}^{\Gamma 3}.
$$
At finite $H$,
$T_{\rm Q}(H)$ shows some increase except for the $\Gamma_3$-type
 with $\mib{H}\|(111)$.
The rate of increase $R\equiv\{\mbox{max}[T_{\rm Q}(H)]-T_{\rm
Q}(0)\}/T_{\rm Q}(0)$
satisfies $R^{\Gamma 3} <R^{\Gamma 5} < R^{\Gamma 3 + \Gamma 5}$.
That is to say, the rate becomes larger
 as the number of components of interaction increases.
%

The types of order parameters on the phase boundary are summarized  in
Table I.
A cusp in the phase diagram means the change from $\langle {O_2}^0
\rangle$ to $\langle {O_2}^2 \rangle$ with increasing $H$.
The magnetic field $H_{\rm c}$ at the cusp
is also shown in Table I.
The phase boundary for the $(\Gamma_3+\Gamma_5)$-type interaction in
weak fields
 has very small dependence on the orientation of magnetic field.
Although $T_{\rm Q}(H)$ in weak fields seems to have almost linear
field-dependence in Figure 1,
it is in fact perpendicular to the $T$-axis at $H=0$.
Since we are interested mainly in the vicinity of the phase boundary
 between the phases I and II,
study of the order parameters at $T<T_{\rm Q}(H)$ is left for future
study.

\subsection{Specific heat and entropy}
Figure 2 shows the  specific heat for the case of the $\Gamma_5$-type
interaction with $\mib H\|(111)$
above the transition temperature $T_{\rm Q}(H)$.
We notice that the MFA gives zero specific heat in zero field
above $T_{\rm Q}(H)$.
It is known in the SA for the Ising model that
the specific heat has a cusp at the transition without a jump
\cite{Brout60}.
We expect a similar behavior in our model although we have not
calculated  the specific heat below the transition temperature.
At weak magnetic field the specific heat decreases monotonically as
temperature increases above $T_{\rm Q}(H)$.
It is evident that the specific heat near $T_{\rm Q}(H)$
becomes sharper and larger with increasing magnetic field.
With further increase of the field beyond $\mu_{\rm B}H \sim 5 J'$,
the peak becomes smaller and the Schottky type anomaly appears.
This is due to the Zeeman splitting of the $\Gamma_8$ levels.
We discuss implication of the result to experiment in \S
\ref{section:discussion}.

We calculate the entropy relative to the value at the transition
temperature by numerical integration of the specific heat.
Figure 3 shows the result.
In the high temperature limit the entropy should tend to $\ln 4$.
Hence we can roughly estimate the value at $T_{\rm Q}(H=0)$ since it
almost saturates in the high temperature end of the calculated range.
It is remarkable that the entropy at the transition temperature is
much less than the full value $\ln 4$ even in the absence of Zeeman
splitting.
We will report the details of thermodynamic quantities including
magnetization for various direction of $\mib H$ and different types of
interactions in a separate paper.

\section{Mechanism of Increased Transition Temperature by Magnetic
Field}
\label{weak field analysis}

\subsection{Decreasing number of fluctuating components}
Our results show pronounced difference of the phase diagram and
entropy between the MFA and the SA.
The simplest model we take proves to be a convenient model to
investigate the effect  of the fluctuations.
In this section we discuss how a magnetic field suppresses the
fluctuation.

When different parameters  are competing for the stability,
 the transition temperature should be lowered in general.
If there are
 many competing order parameters,
 different types of short-range order should occur and
 disturb each other.
Therefore the transition temperature should become lower than
that of a system in which only one of them is favored.
For our model,
the $\Gamma_3$-type interaction has two competing components,
the $\Gamma_5$-type has three and the $(\Gamma_3+\Gamma_5)$-type has
five.
Then, $T_{\rm Q}(H=0)$ in the SA decreases
 with increasing number of components.
The reason is that
 a model with many components has many types of competing orderings.
On the other hand, the MFA cannot deal with this competition and
 gives the same $T_{\rm Q}(H=0)$ for all types of interaction.

This type of competition should be suppressed
with increasing $H$ in our model.
In other words,
 the ordering of the one (or two) of several components becomes
 more favorable than the others, and
the number of competing orderings decreases.
Consequently $T_{\rm Q}(H)$ should increase.
This effect becomes larger as the number of components increases
because the situation changes more drastically
when one of a larger number of components becomes favorable.
Therefore,
this mechanism naturally explains the result that
the rate $R$ of increase in $T_{\rm Q}(H)$
satisfies $R^{\Gamma 3} <R^{\Gamma 5} < R^{\Gamma 3 + \Gamma 5}$.
On the contrary,
we obtained monotonically decreasing $T_{\rm Q}(H)$
for the $\Gamma_3$-type interaction with $\mib{H}\|(111)$.
In this case, such suppression of fluctuations does not occur
because the ordering of ${O_2}^2$
 is as favorable as that of ${O_2}^0$
 owing to the trigonal symmetry.
In order to confirm this mechanism,
we calculated $T_{\rm Q}(H)$ in another model
 which has only one component like the Ising model.
This model does not have such mechanism of suppression.
We indeed find that  $T_{\rm Q}(H)$ does not increase with $H$.

\subsection{Effects of fluctuating magnetic field on spin ordering}
\label{sec:gauss}

The SA  applied to
the Heisenberg model and the XY model does not give
 increasing transition temperature with increasing magnetic field.
One difference
 between the Heisenberg model and  the $\Gamma_5$-type interaction
model
 is that
while only one component among three components
 becomes favorable at finite magnetic fields
  in the latter,
 two components perpendicular to the magnetic field
  become favorable in the former.
As a result, the suppression in the Heisenberg model is rather small.
However, this remark cannot be applied in comparing
the XY model and  the $\Gamma_3$-type interaction.
Another difference, which is probably more important, is that
 a magnetic field is not conjugate to quadrupole moments
 and is coupled with them in a complex way.
To analyze this, it is convenient to use
the pseudo-spin representation of the Zeeman term with
$\mib{H}\|(001)$,
namely, $\{ \sigma^z + (4/7) \tau^z\sigma^z \} \mu_{\rm B}H$.
The quadrupole moment
$\tau^z$ (${O_2}^0$) couples to a kind of
fluctuating field $\sigma^z$ which has the Ising-type distribution.
As shown below, the fluctuating field $\sigma^z$ has a character of
 suppressing fluctuation of quadrupole moments.
However, the Zeeman term plays a dual role of favoring a particular
type among competing  fluctuating fields, and
also destroying the quadrupolar ordering as in antiferromagnets.
In the following, we first take a spin model to
clarify  effects of a fluctuating magnetic field.
Then, we investigate how a static magnetic field affects the
fluctuation in the $\Gamma_3$-type interaction model.
It is shown that the dominant effect of the Zeeman term at weak field
is suppressing quadrupolar fluctuations.

First of all, we take a single classical spin coupled with
a Gaussian fluctuating field.
The Gaussian identity,
\begin{equation}
\label{eq:gaussian}
\int_{-\infty}^{\infty} \frac{\beta{\rm d}\varphi^z}{\sqrt{2 \pi \beta
V^z}}
 \exp[-\beta\frac{(\varphi^z)^2}{2 V^z}
  +\beta \varphi^z S^z]
= \exp[\beta \frac{V^z}{2} (S^z)^2] ,
\end{equation}
shows that a Gaussian fluctuating field $\varphi^z$
 coupled to a classical variable $S^z$
is equivalent to a field $V^z/2$ coupled to $(S^z)^2$.
That is to say,
$\langle (S^z)^2 \rangle$ increases with increasing $V^z$
and the relation
 $-2 \,\partial\Omega/\partial V^z=\langle (S^z)^2 \rangle $
is satisfied where $\Omega$ is the thermodynamic potential.
Therefore,
 the fluctuating field increases the susceptibility
$\chi^{z}=\beta ( \langle (S^z)^2 \rangle- \langle S^z\rangle^2 )$.
On the contrary, $\chi^{x}$ and $\chi^{y}$ are decreased
because of the identity $\langle (S^x)^2+(S^y)^2+(S^z)^2 \rangle
=S(S+1)$.
Note that
the increase of $\chi^{z}$, and the decrease
of $\chi^{x}$ and $\chi^{y}$
are of $O(V^z)$
because $V^z$ is a field conjugate to $(S^z)^2$.
Even for a quantum spin, the discussion above is still valid
provided that $S^z$ is replaced by
 $\beta^{-1}\int_0^\beta {\rm d}\tau S^z(\tau)$.
Namely, the susceptibility $\chi^{z}$ is given by
\begin{eqnarray}
\chi^{z}+\beta \langle S^z\rangle^2 & = & \beta \langle \left\{
\beta^{-1}\int_0^\beta {\rm d} \tau S^z(\tau) \right\}^2 \rangle \\
& = & \int_0^\beta {\rm d}\tau \langle S^z(\tau) S^z(0) \rangle .
\end{eqnarray}
A difference from a classical spin is that
 the susceptibilities are not directly determined by
 the identity $\langle (S^x)^2+(S^y)^2+(S^z)^2 \rangle =S(S+1)$.
For the case of $S=1/2$,
the susceptibilities
in the presence of the Gaussian fluctuating field
with variance $V^z/\beta$
are calculated as
\begin{equation}
\chi^{z}= \frac \beta 4 ,
\;\;
\chi^{x}= \frac\beta 4
           \frac{F(1/2,3/2;\beta V^z/2)}
           {F(1/2,1/2;\beta V^z/2)}   .
\end{equation}
$\chi^{x}$ has the same form as the quotient in the second term
 of eq.(\ref{eq:chiL}) and is a monotonically decreasing function of
$V^z$.
Therefore, the fluctuating field
 decreases $\chi^{x}$ and $\chi^{y}$,
but does not affect $\chi^{z}$.
The reason for the latter is that the Hamiltonian and $S^z$ are
commutative,
and that one has $\left\{
\beta^{-1}\int_0^\beta {\rm d} \tau S^z(\tau) \right\}^2=(S^z)^2= 1/4$
independent of $V^z$.
For $S>1/2$,
$(S^z)^2$ is not a constant
 and thus  $\chi^{z}$ is increased by the fluctuating field.
Note that in any case
the change of susceptibilities is of first order in $V^z$
by the same reason as in the classical system.

Next, we treat the lattice systems.
Let us take the following model as an example which have
 both the Heisenberg-type exchange interaction
and Gaussian fluctuating fields:
\begin{equation}
\label{sigmaSmodel}
{\cal H}_{\varphi S}= - \frac12\sum_{ij} J_{ij} {\mib S}_i\cdot{\mib
S}_j
          -  \sum_i \varphi^z_i S_i^z.
\end{equation}
For classical spin systems and quantum systems with $S>1/2$,
 the local susceptibility $\chi_{\rm L}^z$ at each site
is increased by the Gaussian fluctuating field
 as well as that of the single spin.
Consequently,
the transition temperature to the ordered state should increase.
In the mean-field approximation,
the susceptibility $\chi_{\mibs Q}=\chi_{\rm L}/(1-J_{\mibs
Q}\chi_{\rm L})$
 for a certain wave number $\mib{Q}$
 diverges at the phase transition.
Hence the transition temperature increases
if $\chi_{\rm L}^z$ is increased by the fluctuating field
$\varphi^z_i$.
Even in quantum spin systems with $S=1/2$,
 the transition temperature should still
increase by the fluctuation effect.
As $V^z$ increases,
ordering of $S^x$ or $S^y$ should become less favorable.
Therefore, in the limit of large $V^z$, only the $S^z$ ordering
 should be possible.
This means that the fluctuation among components is suppressed.
Certainly, the transition temperature increases with $V^z$
 in the SA to this model.
Further, the local susceptibility $\chi^{z}_{\rm L}$ increases here.
The reason is that $V^z$ makes $\chi_{\rm L}^x$ and $\chi_{\rm L}^y$
decrease.
Then, $\tilde{J}^x$ and $\tilde{J}^y$ decrease and
 $\chi_{\rm L}^z$ increases.
On the other hand,
when we solve the model by the mean-field approximation,
the transition temperature $T_{\rm c}$ for the $S^x$ and $S^y$
orderings
decreases with increasing $V^z$,
while $T_{\rm c}$ for the $S^z$ ordering does not change.
Namely, in the spin model defined by
eq.(\ref{sigmaSmodel}) with $S=1/2$
the transition temperature does increase with field
in contrast to the result of the mean field approximation.

This fluctuating field has similar effect
even if
 $\varphi^z$ is not a Gaussian-distributed variable
 but the Ising-type variable taking values of
 $\varphi^z=\pm \sqrt{V^z/\beta}$.
This field resembles $h \sigma^z$
in the Zeeman term with $H\|(001)$
represented by the pseudo-spins for the quadrupolar model.
We discuss it in the next subsection.

\subsection{Two competing effects of magnetic field on the quadrupolar
ordering}

\label{sec:G3xy}

In the pseudo-spin representation,
the interaction Hamiltonian for the $\Gamma_3$-type interaction
 is the same as that of the XY model.
The only difference between them is the Zeeman term.
Nevertheless, their phase diagrams by the SA with $\mib{H}\|(001)$
 show a remarkable difference;
while the transition temperature increases with $H$
 in the $\Gamma_3$-type interaction,
it does not in the XY model.
In this subsection we discuss the reason for
 the difference in weak fields.

The Hamiltonian of the $\Gamma_3$-type interaction with
$\mib{H}\|(001)$ is written as
\begin{eqnarray}
{\cal H} & = &
 - \frac12\sum_{ij} J_{ij}(\tau^x_i\tau^x_j+\tau^z_i\tau^z_j)
   -  \sum_i \mu_{\rm B}H \sigma^z_i\left(1+\frac47\tau^z_i\right) ,
\nonumber \\ \label{Boriginal} & &
\\
 & \equiv & {\cal H}_{\rm int}
   -  \sum_i   h \sigma^z_i\left(1+\frac47\tau^z_i\right),
\end{eqnarray}
where $h$ refers to $\mu_{\rm B}H$.
In this expression, the $\mib\sigma$-spin emerges only as $\sigma^z$.
Hence,
 one can treat it as a classical variable taking values $\pm1$.
To study the structure of the partition function,
 we introduce path-integral representation for the $\mib\tau$-spin.
Since the explicit expression for the Berry phase term is
 not necessary here, we write it symbolically
 as ${\cal L}_{\rm B}$.
\begin{equation}
Z = \sum_{\{\sigma^z_i\}} \int {\cal D}{\mib \tau}
  \exp\left[
  -\int_0^{\beta} {\rm d}\tau \{ {\cal L}_{\rm B}(\tau)+{\cal
H}(\tau)\}\right] ,
\end{equation}
After summation over $\sigma^z$,
we neglect terms of higher order in $\beta h$
 to investigate the behavior in weak fields.
We use the approximation $\cosh x \simeq e^{x^2/2}$ for $x\ll 1$.
Here we have
$$
x = h \int_0^{\beta} \left(1+\frac47\tau^z_i(\tau)\right) {\rm d}\tau.
$$
Then we use the identity eq.(\ref{eq:gaussian}) to introduce the
fluctuating field $\varphi^z_i$.  Finally we
turn to the operator representation with Tr instead of $\int {\cal
D}{\mib \tau}$. The calculation explained above goes as
\fulltext
\begin{eqnarray}
Z & \simeq & 2^N \int {\cal D}{\mib \tau}
  \exp\Big[
 -\int_0^{\beta}\{{\cal L}_{\rm B}(\tau)+{\cal H}_{\rm
int}(\tau)\}{\rm d}\tau
 + \frac{(\beta h)^2}2 \sum_i
\left\{ \frac1 \beta
\int_0^{\beta} \left(1+\frac47\tau^z_i(\tau)\right) {\rm d}\tau\right\}^2
      \Big]
\\
 & = & 2^N \int {\cal D}{\mib \tau}
\prod_i\int_{-\infty}^{\infty} \frac{{\rm d}\varphi^z_i}{\sqrt{2 \pi h^2}}
 \exp\Big[
  -\int_0^{\beta}\{{\cal L}_{\rm B}(\tau)+{\cal H}_{\rm
int}(\tau)\}{\rm d}\tau
\nonumber\\ & & \makebox[13em]{}
+\sum_i \left\{ -\frac{(\varphi^z_i)^2}{2 h^2}
 +  \varphi^z_i \int_0^{\beta} \left(1+\frac47\tau^z_i(\tau)\right)
{\rm d}\tau
    \right\}  \Big]
\\
 & = & 2^N
\prod_i\int_{-\infty}^{\infty} \frac{{\rm d}\varphi^z_i}{\sqrt{2 \pi h^2}}
\exp\left[-\sum_i \frac{(\varphi^z_i)^2}{2 h^2}\right]
{\rm Tr}
 \exp\Big[
  - \beta\left\{ {\cal H}_{\rm int}
 -\sum_i   \varphi^z_i \left(1+\frac47\tau^z_i\right)
      \right\}\Big]
\label{B6}
\\
 & = & 2^N
\prod_i\int_{-\infty}^{\infty} \frac{7 {\rm d}\varphi^z_i}{4\sqrt{2 \pi
h^2}}
\exp\left[\sum_i \left( - \frac{(\varphi^z_i)^2}{2 (4h/7)^2}
   +\frac12\beta^2 h^2\right)\right]
\nonumber\\ & &
\makebox[8em]{}\times {\rm Tr}
 \exp\Big[
  - \beta\left\{ {\cal H}_{\rm int}
  - \sum_i\left( \varphi^z_i \tau^z_i + \frac47\beta h^2 \tau^z_i
\right)
      \right\}\Big] . \label{B7}
\end{eqnarray}
\halftext
In the expression (\ref{B6}),
the only difference from the original expression (\ref{Boriginal})
is that the Ising-type distribution $h\sigma^z_i$ is
 replaced by $\varphi^z_i$,
which obeys the Gaussian distribution with the variance $h^2$.
Then,
we replaced $\varphi^z_i$ by $(7/4) \varphi^z_i + \beta h^2$
 in deriving eq.(\ref{B7}).

The linear term with respect to $h$ is absent in
eq.(\ref{B7}) and
$\tau^z$ couples to $h^2$ because of the time reversal symmetry.
The term
$-(4/7) \beta h^2\tau^z_i$ is of
the same form as the Zeeman term in the XY model
and thus it works
to induce $\langle\tau^z_i\rangle$ uniformly
and to destroy the antiferro-type ordering.
Since the lowest order contribution of a magnetic field
 to the susceptibilities in the XY model is the square of the field,
the lowest-order contribution of the term $-(4/7) \beta h^2\tau^z_i$
to the strain susceptibilities should be the square of
$-(4/7) \beta h^2$, namely $O[(\beta h)^4]$.

On the other hand, the Gaussian fluctuating field $\varphi^z_i$
with variance $(4h/7)^2$
affect the susceptibilities by $O[(\beta h)^2]$
 as described in the previous subsection.
Therefore, it should make a dominant contribution at weak fields.
This term makes the ordering of $\tau^z$ favorable.
After all  a weak magnetic field suppresses the fluctuation
and increases the transition temperature.

In the previous paper of Uimin and the present authors,\cite{UKF}
the pseudo-spins for the quadrupole moments
are approximated by classical spins.
In the previous subsection,
we have shown that
 the fluctuating field $\varphi_i^z$ has smaller influence on
$\chi^{z}$
in the quantum spin with $S=1/2$ than that in the classical spin.
Hence the approximation in ref.\citen{UKF} enhances the effect
of the fluctuating field $\varphi_i^z$ in eq.(\ref{B7}).
The classical treatment should be the reason why
the result of the $\Gamma_3$-type interaction with $\mib{H}\|(001)$
shows much larger increase of $T_{\rm Q}(H)$ than that of the present
paper.

\begin{tabular}{c}
\hline
\end{tabular}

\newpage

\section{Discussion and Concluding Remarks}
\label{section:discussion}

In this paper we have shown for quadrupolar interaction systems that
a magnetic field has two contrasting effects.
One effect is inducing quadrupole moments uniformly
and destroying the antiferro-quadrupolar ordering.
The induced quadrupole moments are even functions of magnetic field
owing to the time reversal symmetry,
and the change of $T_{\rm Q}(0)$ is very small ($\propto H^4$) at weak
fields.
The other effect is suppressing the competition
among different components of order parameters.
Since the resultant change is of $O(H^2)$,  the latter effect
dominates at weak fields.
Hence the transition temperature increases as $H$ increases.

For comparison with experimental results in CeB$_6$,
a crude estimate of $J'$ is obtained by comparing the theoretical
value of  $T_{\rm Q}(0)$ with the experimental one.
With the value of $J'$ so determined,
the unit for magnetic field is also fixed.
Let us assume for the moment  the $\Gamma_5$ order.
Since the numerical value of
 $T_{\rm Q}(0)$ in units of $J'/k_{\rm B}$
 is the same order as the experimental value in units of Kelvin,
 one may roughly estimate  $J'/k_{\rm B} \sim 1 {\rm K}$,
 and the unit for magnetic field as 1.5 T (tesla).
%
Then the increase of $T_{\rm Q}(H)$ in Figure 1 turns out not so
strong as the experimentally observed one,
and the re-entrant field is much smaller than the actual one which is
larger than the maximum of available magnetic field.

On the other hand, results of the specific heat at  weak fields agree
qualitatively with experimental ones both for the magnitude and the
temperature dependence.
We emphasize that the lowering of the cubic symmetry is not necessary
to explain the entropy smaller than $\ln 4$ at the transition
temperature, and
the large tail of the specific heat experimentally observed.
The temperature range of a weak Schottky type anomaly found in Figure
2 for $H \sim$ 12 T and larger overlaps actually with the experimental
$T_{\rm Q}(H)$ up to 15 T in CeB$_6$ \cite{Sera}.
Hence the anomaly is not seen experimentally.
However in the alloy system Ce$_{0.5}$La$_{0.5}$B$_6$, such Schottky
anomaly is indeed observed \cite{Sera2} due to the lowered $T_{\rm
Q}(H)$.

For serious comparison with experimental results, we have to consider
two directions of development on the basis of the present study.
The first is to choose a realistic model to describe the interaction
among quadrupole moments at different sites within the SA.
Concerning the intersite interaction, the role of the range \cite{UKF}
and types including higher multipoles \cite{Ohkawa1,Ohkawa2,Shiina}
should be reexamined with inclusion of fluctuations.
With the intersite octupolar interaction, the effect of magnetic field
on the shift of $T_{\rm Q}(H)$ becomes larger and should compare more
favorably with experiment.
Although we have neglected the higher crystal field levels other than
the ground level,
the crystal field splitting is so small in TmTe that the $\Gamma_6$
and the $\Gamma_7$
doublets seem to be within a few meV above the ground $\Gamma_8$
level.
 Hence, it is interesting to investigate how the crystal field
splitting
affects the phase diagram of the quadrupolar ordering.

Another direction of development is including dynamical and quantum
effects of fluctuations.
The most important may be the Kondo effect in the case of CeB$_6$ but
not
in the case of insulating systems such as TmTe.  In the latter case
the quantum fluctuation of multipole moments should persist even
without the Kondo effect \cite{KF}.
In order to deal with dynamical effects in the effective medium
theory, numerical methods such as the resolvent method, the quantum
Monte Carlo or the numerical
renormalization group seem promising.
We hope to report on these development in the near future.

\acknowledgement
The authors  thank G. Uimin for discussion about
the fluctuation among orderings of different wave numbers.
Thanks are also due to T. Matsumura, T. Suzuki and M. Sera for
helpful conversations on experimental results.

\begin{fulltable}
\caption{Order parameters on the phase boundary.
  Here $H_{\rm c}$ refers to the critical magnetic field
at which the type of order parameters changes.}
\label{orderpara}
\begin{fulltabular}{c|ccc|c}
   & \multicolumn{2}{c}{$\mib{H}\|(001)$} & $\mib{H}\|(111)$  & \\
   &  $H < H_{\rm c}$  &  $H > H_{\rm c}$ & & $H_c$\\
\cline{1-2}\cline{2-5}
 $\Gamma_3$
& $\langle {O_2}^0 \rangle$
&
  $\langle {O_2}^2 \rangle$
& $\langle {O_2}^0 \rangle$ or $\langle {O_2}^2 \rangle$
& 5.7 \\
%
 $\Gamma_5$
& \multicolumn{2}{c}{$\langle O_{xy} \rangle$}
& $\langle O_{yz}+O_{zx}+O_{xy}\rangle/\sqrt{3}$
& --- \\
%
 $\Gamma_3+\Gamma_5$
& $\langle {O_2}^0 \rangle$
& $\langle {O_2}^2 \rangle$ or $\langle O_{xy} \rangle$
& $\langle O_{yz}+O_{zx}+O_{xy}\rangle/\sqrt{3}$
& 4.8 \\
\end{fulltabular}
\end{fulltable}

\begin{fullfigure}
\epsfxsize = 9cm \epsfbox{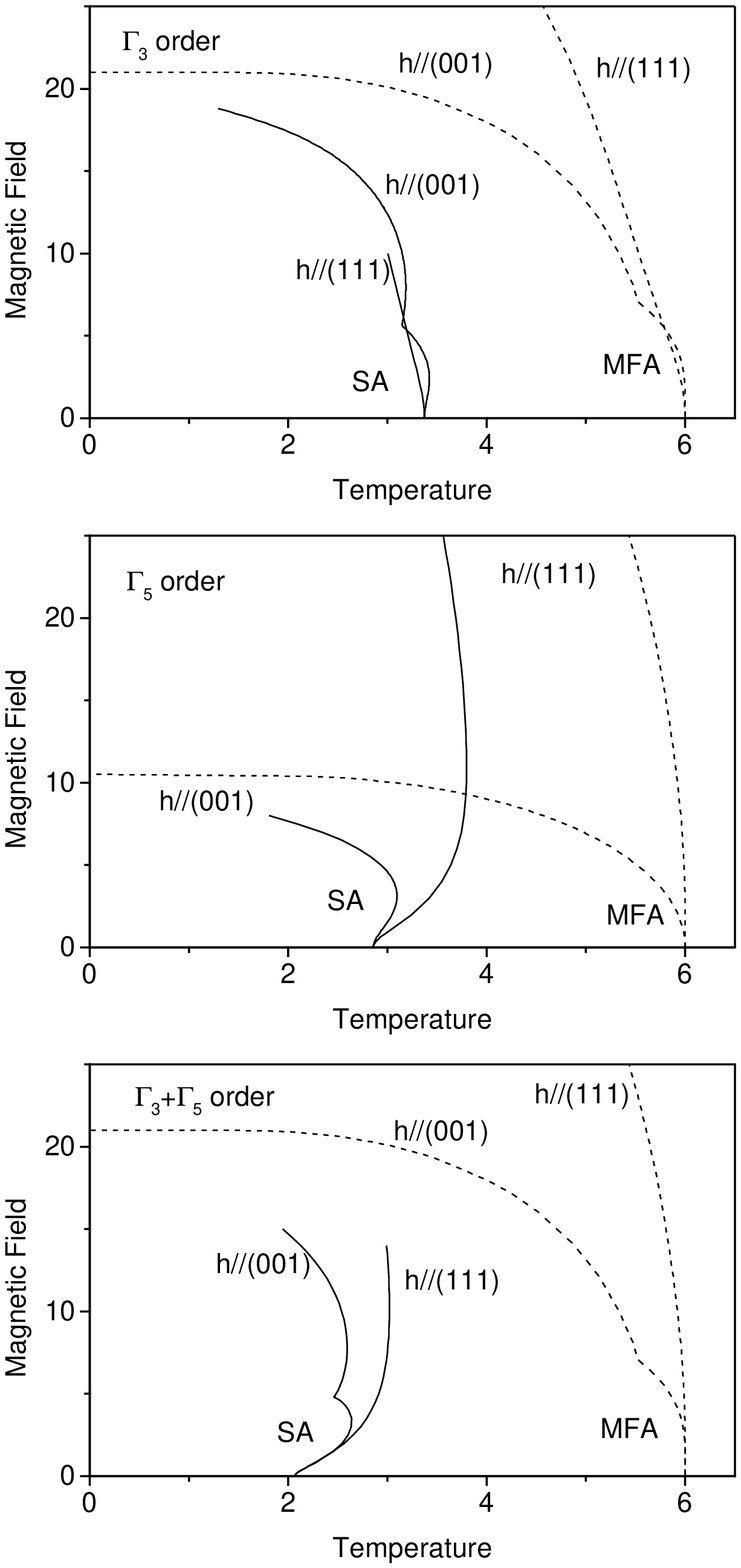}
\caption{The phase diagrams obtained
 by the SA(solid lines) and the MFA(broken lines).}
\end{fullfigure}

\begin{fullfigure}
\epsfxsize = 11cm \epsfbox{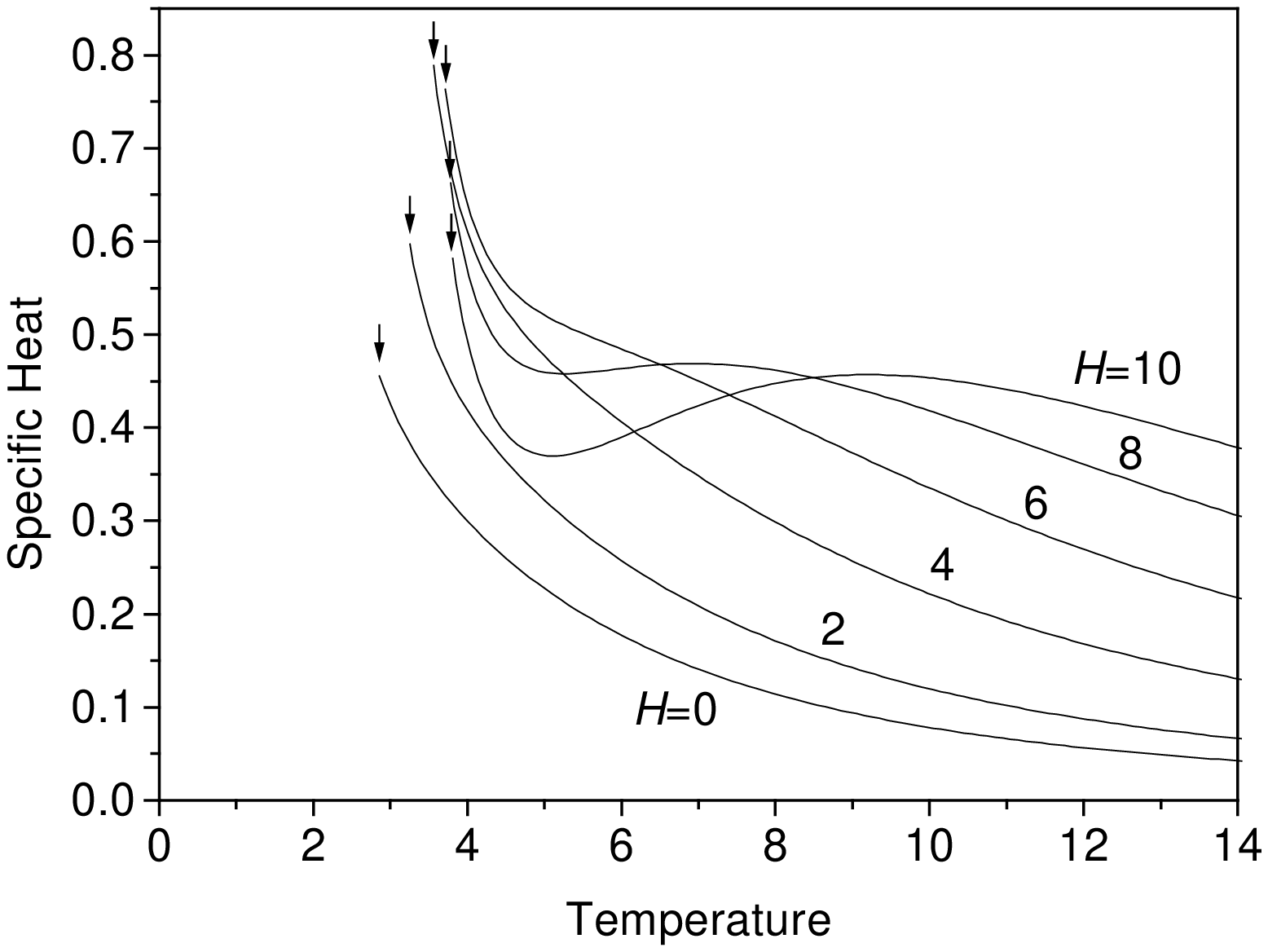}
\caption{The specific heat above the transition temperature
 obtained by the present theory
 for the $\Gamma_5$-type interaction with $\mib H\|(111)$.
Each transition temperature is indicated by an arrow.
The unit of the ordinate corresponds to
8.31J/(mol$\cdot$ K).}
\end{fullfigure}

\begin{fullfigure}
\epsfxsize = 11cm \epsfbox{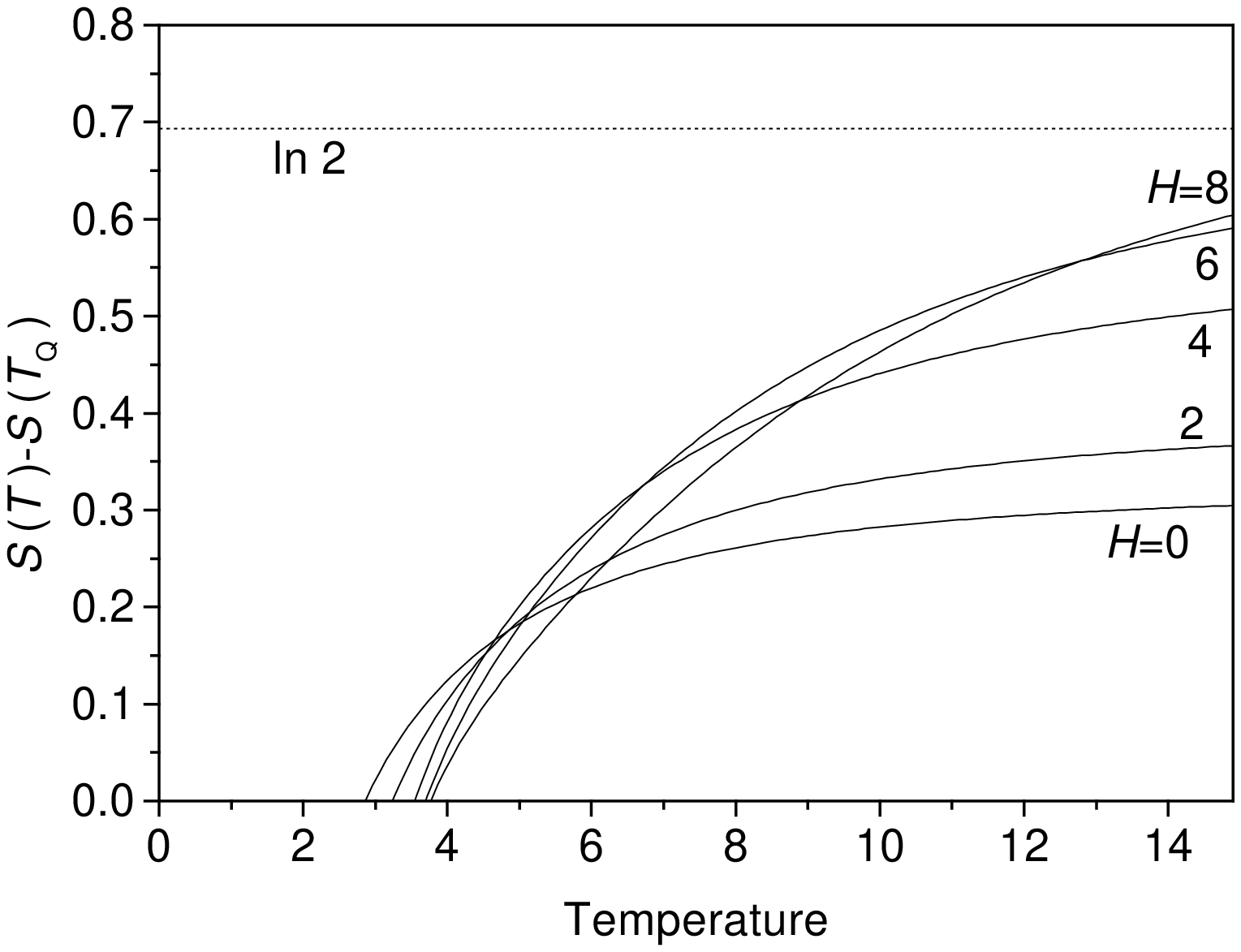}
\caption{The change of entropy relative to the value at the transition
temperature.  }
\end{fullfigure}

\end{document}